\documentstyle[aps,preprint,epsf,amssymb]{revtex} 
\def\ldss{\left\langle\kern -3pt\left\langle}
\def\rdss{\right\rangle\kern -3pt\right\rangle}
\def\lds{\left\langle\kern -4pt\left\langle}
\def\rds{\right\rangle\kern -4pt\right\rangle}
\def\ldb{\left\langle\kern -7pt\left\langle}
\def\rdb{\right\rangle\kern -7pt\right\rangle}

\begin{document}
\draft
\title{Dynamics of relaxor ferroelectrics}
\author{R. Pirc, R. Blinc, and V. Bobnar}
\address{Jo\v zef Stefan Institute, P.O. Box 3000, 1001 Ljubljana, Slovenia}
\date\today
\maketitle

\tighten
\begin{abstract}                    

We study a dynamic model of relaxor ferroelectrics based on the spherical 
random-bond---random-field model and the Langevin equations of 
motion written in the representation of eigenstates of the random
interaction matrix. The solution to these equations is obtained in the
long-time limit where the system reaches an equilibrium state in the
presence of random local electric fields. The complex dynamic linear and 
third-order nonlinear susceptibilities $\chi_1(\omega)$ and $\chi_3(\omega)$,
respectively, are calculated as functions of frequency and temperature.
In analogy with the static case, the dynamic model predicts a narrow 
frequency dependent peak in $\chi_3(T,\omega)$, which mimics a transition 
into a glass-like state, but a real transition never occurs in the case 
of non-zero random fields. A freezing transition can be described 
by introducing the empirical Vogel-Fulcher (VF) behavior of the relaxation time 
$\tau$ in the equations of motion, with the VF temperature $T_0$ 
playing the role of the freezing temperature $T_f$. The scaled third-order 
nonlinear susceptibility $a_3'(T,\omega) = \overline{\chi}_3'(\omega)
/\overline{\chi}_1'(3\omega)\overline{\chi}_1'(\omega)^3$, where the bar  
denotes a statistical average over $T_0$, shows a crossover
from paraelectric-like to glass-like behavior in the quasistatic regime 
above $T_f$. The shape of $\overline{\chi}_1(\omega)$ and 
$\overline{\chi}_3(\omega)$---and thus of $a_3'(T,\omega)$---depends crucially 
on the probability distribution of $\tau$. It is shown that for a linear 
distribution of VF temperatures $T_0$, $a_3'(T,\omega)$ has a peak near 
$T_f$ and shows a strong frequency dispersion in the low temperature region.

\end{abstract}

\pacs{PACS numbers:77.84.Dy, 64.70.Pf, 77.22.-d}
\narrowtext

\newpage
\section{Introduction}

Relaxor ferroelectrics (or relaxors) represent a new low-temperature state of 
polar dielectrics, which can be regarded as an intermediate state between 
dipolar glasses and normal ferroelectrics \cite{C1,VR1}. Some of the concepts 
developed for dipolar glasses, such as the Edwards-Anderson (EA) order parameter,
are applicable to relaxors as well, as recently shown for 
PbMg$_{1/3}$Nb$_{2/4}$O$_3$ (PMN) \cite{B1}, PbSc$_{1/2}$Ta$_{1/2}$O$_3$ (PST) 
\cite{B4}, and Pb$_{1-x}$La$_{2x/3}$Zr$_y$Ti$_{1-y}$O$_3$ (PLZT) \cite{B5}.
In contrast to dipolar glasses, where elementary dipole moments exist on
the atomic scale, the relaxor state is characterized by the presence of 
nanosized polar clusters of variable sizes. This picture constitutes the basis
of the superparaelectric model \cite{C1} and of the more
recent reorientable polar cluster model of relaxors \cite{VR1,VR2}.
By including  explicitly the long-range frustrated intercluster interactions
of a spin glass type into this picture, one arrives at the so-called spherical 
random-bond---random-field (SRBRF) model of relaxor ferroelectrics \cite{PB}, 
which is capable of describing the static behavior of relaxors, such as the 
line shape of quadrupole perturbed NMR in PMN \cite{B1} and PST \cite{B4}, 
and the sharp increase of the quasistatic third-order nonlinear dielectric 
constant \cite{B1,B5}. 

The unusually large value of the static linear dielectric
permittivity can also be explained within the framework of the SRBRF model 
if one assumes that the mean value of the random coupling $J_0$ is very
close to---but slightly smaller than---its r.m.s. variance $J$, whereas in
dipolar glasses the latter is usually dominant. By including an {\it ad hoc}
electric field dependence of $J_0$ into the model, one can furthermore describe 
the transition from the relaxor to an inhomogeneous ferroelectric state for 
fields $E$ exceeding a critical value $E_c$ \cite{B6}. It should be noted,
however, that the random electric fields, which exist both in dipolar 
glasses \cite {PTB} and in relaxors \cite{VR1,VR2,W1}, seem to be much weaker 
in the latter case. It is interesting to note that Korner et. al \cite{Korner}
have reported a relaxor phase in the dipolar glass DRADP in a narrow range
of concentration just above the dipolar glass phase, where the system is
expected to behave as an inhomogeneous antiferroelectric. 

While the static SRBRF model describes a relaxor system in thermodynamic 
equilibrium, there are a number of phenomena suggesting that relaxors,
in particular their low temperature state, are dominated by
nonequilibrium effects. Typical examples are the difference between 
the field-cooled and zero-field cooled static dielectric constant,
and the occurrence of strong frequency dispersion in both the linear
and nonlinear dielectric permittivity at low temperatures. Is is
clear that these properties can only be discussed within a dynamic
model. In the present paper, we introduce a dynamic model, which an
extension of the SRBRF model to dynamic problems. Following Vugmeister
and Rabitz \cite{VR1,VR2} we assume that polar clusters can reorient with 
a characteristic relaxation time $\tau$ and write down the corresponding
Langevin equations of motion, which are based on the static SRBRF 
Hamiltonian. These equations explicitly contain the frustrated 
interactions between the polar clusters and thus allow us to study 
the effects of these interactions on both the equilibrium and 
nonequilibrium properties. In particular, we will discuss here the
anomalous temperature dependence of the nonlinear dielectric 
susceptibility and the crossover from the paraelectric-like to
inhomogeneous ferroelectric-like behavior observed in PMN and 
PLZT \cite{B5}. As in spin glasses, the Langevin equations based on
the spherical model can be solved exactly \cite{CD1}; some additional
features appear in view of the presence of random fields \cite{CD2}. 
Here we will focus on the asymptotic solutions corresponding to 
equilibirum dynamics, such as observed in a typical dielectric relaxation
experiment. 

In Sec. II we introduce the uniaxial SRBRF model Hamiltonian in the
representation of eigenstates of the random interaction and write down the
Langevin equations of motion. The asymptotic solution is studied in Sec. 
III, where the static linear and nonlinear susceptibilities are derived.
In Sec. IV the dynamic linear response is given, and in Sec. V
the corresponding results for the third-order nonlinear response
are derived. Finally, in Sec. VI we present our conclusions.

\section{Dynamic SRBRF model}

In general, the polarization of $i$-th polar cluster, $i = 1,2,...,N$, is
a three component $(n=3)$ vector $\vec{S}_i = (S_{ix},S_{iy},S_{iz})$, 
its length being restricted solely by the spherical condition 
$\sum_i (\vec{S}_i)^2 = 3N$. In the present work we will discuss the simpler
uniaxial $(n=1)$ case $-\sqrt{N} < S_i < +\sqrt{N}$, where $S_i$ is subject
to the spherical condition
\begin{equation}
\sum_{i=1}^N S_i^2 = N \; .
\label{sc}
\end{equation}
The SRBRF model Hamiltonian is thus
\begin{equation}
{\cal H} = -{1\over 2}\sum_{ij}J_{ij}S_iS_j - \sum_ih_iS_i - gE\sum_iS_i \; ,
\label{H}
\end{equation}
where $J_{ij}$ are the randomly frustrated intercluster interactions,
$h_i$ local random electric fields, $E$ an applied uniform electric field, 
and $g$ the appropriate dipole moment \cite{PB}. As usual, $J_{ij}$ is
assumed to be infinitely ranged and distributed according to Gaussian 
statistics with average value $J_0/N$ and cumulant variance $J^2/N$. The
Gaussian random fields $h_i$ are characterized by the random average
\begin{equation}
[h_ih_j]_{av} = \Delta \delta_{ij} \; .
\label{hi}
\end{equation}
The uniaxial SRBRF model (\ref{H}) has potential applicability to uniaxial 
relaxors such as Sr$_{1-x}$Ba$_x$Nb$_2$O$_6$ (SBN). The present results can 
be, however, generalized to the isotropic $n=3$ case as long as
there is no mixing of the $x,y,z$ components \cite{PB}. 

The Langevin equations of motion for the variables $S_i(t)$ are written as
\begin{equation}
\tau {\partial S_i(t)\over \partial t} = - {\partial (\beta{\cal H})
\over \partial S_i} - 2z(t)S_i(t) + \xi_i(t) \; .
\label{em1}
\end{equation}
Here $\tau$ is the characteristic relaxation time for the reorientation of 
polar clusters. Eq.\ (\ref{em1}) implies that $\tau$ is site
independent, however, some variation of $\tau$ across the system should
in principle not be excluded, resulting in a distribution of relaxation 
times \cite{VR1,VR2}. The function  $z(t)$ plays the role of a Langrange multiplier
enforcing the spherical condition (\ref{sc}) at all times \cite{CD1}.

The stochastic Langevin forces $\xi_i(t)$ ensure the proper equilibrium 
distribution and are determined by their ensemble averages
\begin{equation}
\langle \xi_i(t) \xi_j(t')\rangle_{av} = 2\tau \delta_{ij} \delta (t-t') \; .
\label{LF}
\end{equation}

Following the theory of spherical spin glasses we now transform to the 
representation of eigenstates $\psi_{\lambda}(k)$ and eigenvalues $J_{\lambda}$ 
of the random matrix $J_{ij}$ \cite{KTJ,CD1,CD2}. This is done in two steps 
\cite{M1}: First, one introduces "spin wave" states 
$S_k = N^{-1/2}\sum_i \exp(ikR_i)S_i$; next, these are expanded in normal 
modes 
\begin{equation}
S_{\lambda} = \sum_k \psi_{\lambda}(k) S_k \; .
\label{Slambda}
\end{equation}
The transformed equation of motion (\ref{em1}) becomes explicitly
\begin{equation}
{\partial S_{\lambda}\over \partial t} = \beta[J_{\lambda}-2z(t)]S_{\lambda}
+\beta h_{\lambda}S_{\lambda} + g\Psi_{\lambda}(0)E(t)S_{\lambda}
+\xi_{\lambda}(t) \; .
\label{emd}
\end{equation}
Here $\Psi_{\lambda}(0) = N\psi_{\lambda}(0)$ and we have rescaled the time to a 
new dimensionless variable $t \to t/\tau$. Assuming a field $E(t)$ applied at 
$t=0$ and introducing the integrating factor 
\begin{equation}
\phi_{\lambda}(t) = \exp \left[ \beta J_{\lambda}t - 
2\int_0^t dt' z(t')\right] 
\label{phi}
\end{equation}
we obtain the solution
\begin{equation}
S_{\lambda}(t) = \phi_{\lambda}(t)S_{\lambda}(0) + \int_0^t dt_1
{\phi_{\lambda}(t)\over \phi_{\lambda}(t_1)}\left[\beta h_{\lambda}
+\beta g\Psi_{\lambda}(0)E(t_1)+\xi_{\lambda}(t_1)\right] \; .
\label{St}
\end{equation}
The correlation function
\begin{equation}
C(t,t') = {1\over N}\sum_{\lambda} \langle S_{\lambda}(t)S_{\lambda}(t')
\rangle_{av} \;
\label{cf}
\end{equation}
must satisfy the equal time relation $C(t,t) = 1$ at all times in view of
the spherical condition (\ref{sc}). From Eqs.\ (\ref{St})-(\ref{cf})
with the aid of Eq.\ (\ref{hi}) we thus find
\begin{eqnarray}
1 = && \lds \phi_{\lambda}(t)^2S_{\lambda}(0)^2 \rds_{\!0}
  + 2\int_0^t dt_1 \ldb {\phi_{\lambda}(t)^2\over 
 \phi_{\lambda}(t_1)^2}\rdb_{\!0} + \beta^2\Delta\int_0^tdt_1\int_0^tdt_2
\ldb {\phi_{\lambda}(t)^2\over \phi_{\lambda}(t_1)\phi_{\lambda}(t_2)}
\rdb_{\!0} \nonumber\\ 
&&  +\beta^2g^2\int_0^tdt_1\int_0^tdt_2
\ldb {\phi_{\lambda}(t)^2\over \phi_{\lambda}(t_1)\phi_{\lambda}(t_2)}
\rdb E(t_1)E(t_2) \; \; .
\label{ctt}
\end{eqnarray}
This is an implicit equation for the Lagrange multiplier $z(t)$. The two types
of averages are defined as
\begin{equation}
\langle\!\langle f_{\lambda}\rangle\!\rangle_{0} 
\equiv {1\over N}\sum_{\lambda} f_{\lambda} 
= \int dJ_{\lambda} \, \rho_0 (J_{\lambda}) f(J_{\lambda}) \; ;
\label{av0}
\end{equation} 
\begin{equation}
\langle\!\langle f_{\lambda}\rangle\!\rangle
 \equiv {1\over N}\sum_{\lambda} \Psi_{\lambda}(0)^2
f_{\lambda} = \int dJ_{\lambda} \, \rho (J_{\lambda}) f(J_{\lambda}) \; ,
\label{av}
\end{equation}
where $\rho_0(J_{\lambda})$ and $\rho(J_{\lambda})$ are the densities of 
eigenvalues in the $k\ne 0$ and $k=0$ sector of the spectrum, respectively.
The eigenvalues $J_{\lambda}$ have a continuous spectrum $-2J < J_{\lambda} < 2J$.
If $|J_0| > J$, there is also a discrete eigenvalue at $J_m = J_0 + J^2/J_0$
\cite{KTJ}. Here we will only discuss the case $|J_0| < J$. The $k\ne 0$ 
density of states is given by the Wigner semicircle law \cite{KTJ,M1}
\begin{equation}
 \rho_0(J_{\lambda}) = {1\over 2\pi J^2} \left (4J^2 - 
 J_{\lambda}^2\right )^{1/2} \; ,
\label{rho0}
\end{equation}
The $k=0$ sector, on the other hand, has the density \cite{M1}
\begin{equation}
 \rho(J_{\lambda}) = {1\over 2\pi (J^2+J_0^2-J_0J){\lambda})} 
 \left (4J^2 - J_{\lambda}^2\right )^{1/2} \; .
\label{rho}
\end{equation}
This density of states has a statistical weight ${\cal O}(1/N)$ and
is thus relevant only in averages containing factors of the type
$\Psi_{\lambda}(0)^2 \propto {\cal O}(N)$.

The dielectric polarization of the system can be expressed in terms of
the solution (\ref{St}) as
\begin{equation}
P(t) = {1\over N}\sum_{\lambda} g\Psi_{\lambda}(0)\phi_{\lambda}(t)S_{\lambda}(0)
+\beta g^2\int_0^t dt_1 \ldb {\phi_{\lambda}(t)\over
\phi_{\lambda}(t_1)}\rdb E(t_1) \; .
\label{Pt}
\end{equation}

As shown by Cugliandolo and Dean \cite{CD2}, for times larger than a limiting 
time $t_c$ the system in which $\Delta \ne 0$ will always reach an equilibrium 
state and will thus be characterized by equilibrium dynamics. All information
about the initial state $S_{\lambda}(0)$ is lost for $t \gg t_c$, i.e.,
the first term in Eq.\ (\ref{Pt}) becomes irrelevant. In the present
case, $t_c$ is estimated as $t_c \approx 2\tau JT/\Delta$. Typically, the 
asymptotic regime $t \gg t_c$ is explored in a dielectric relaxation experiment. 
In the following, we will limit ourselves to this regime. Also, for 
simplicity we will henceforth set $g=1$.


\section{Static dielectric response}

We first consider the case of a static electric field $E(t) = E$ applied at
$t=0$. At asymptotic times $t/\tau \gg 1$ the system reaches equilibrium
and the Lagrange multiplier $z(t)$ tends to a constant value $z$. Thus the
function (\ref{phi}) becomes
\begin{equation}
\phi_{\lambda}(t)\sim e^{-g_\lambda t} \; ; \;\;\;\;
(g_{\lambda} \equiv 2z - \beta J_{\lambda}) \; , 
\label{phia}
\end{equation}
and we can evaluate the integrals in Eqs.\ (\ref{ctt}) and (\ref{Pt}). 
Assuming that $2z > \beta J_{\lambda}$ for all $\lambda$ 
(to be justified later) we derive the equation for $z$:
\begin{equation}
1 = \ldb {1\over 2z - \beta J_{\lambda}}\rdb_{\!0}
+\beta^2\Delta\ldb {1\over (2z - \beta J_{\lambda})^2}\rdb_{\!0}
+\beta^2E^2\ldb {1\over (2z - \beta J_{\lambda})^2}\rdb \; .
\label{eqz}
\end{equation}

The static linear susceptibility $\chi_1 = (\partial P/\partial E)_{E=0}$ 
is derived from Eqs.\ (\ref{Pt}) and (\ref{phia}):
\begin{equation}
\chi_1 = \beta \ldb {1\over g_{\lambda}}\rdb \; .
\label{chi1s}
\end{equation}
The averages in Eqs.\ (\ref{eqz}) and (\ref{chi1s}) can be expressed
in terms of the generalized averages obtained by adding an imaginary 
generating field $iy$ to the variable $g_{\lambda}$, namely,
\begin{equation}
\chi_1^{[n]}(y) \equiv \beta\ldb {1\over \;\;\; 
(g_{\lambda} - iy)^{(n+1)}}\rdb \; .
\label{chi1[n]}
\end{equation}
These averages can be evaluated with the aid of Eqs.\ (\ref{av0})-(\ref{rho})  
for $n=0$, differentiating $n$ times with respect to $iy$, and setting $y=0$. 
For example, from Eqs.\ (\ref{chi1s}) and (\ref{chi1[n]}) with $n=0$ and 
$y=0$ we find:
\begin{equation}
\chi_1 = \chi_1^{[0]}(0) = {z-r -\beta J_0\over 
\beta (J^2+J_0^2)-2\beta J_0z} \; ,
\label{chi1r}
\end{equation}
where $r \equiv \sqrt{z^2-\beta^2J^2}$. 

The $n=1$ average is given by
\begin{equation}
\chi_1^{[1]}(y) = {\beta\over 2} \; {z(y)-r(y)\over D(y)} 
\left[{1\over r(y)} - {2\beta J_0\over D(y)}\right]
+ {\beta^3 J_0^2\over D(y)^2} \; ,
\label{chi1[1]}
\end{equation}
where $z(y) \equiv z - iy/2$, $r(y) \equiv \sqrt{z(y)^2-\beta^2J^2}$, and
$D(y) \equiv \beta^2(J^2+J_0^2)-2\beta J_0z(y)$.

The above equation for $z$, Eq.\ (\ref{eqz}),  becomes in this notation:
\begin{equation}
1 = {1\over \beta} \chi_1^{[0]}(0)_0 + \beta\Delta \chi_1^{[1]}(0)_0
+\beta E^2 \chi_1^{[1]}(0) \; ,
\label{eqzf}
\end{equation}
where the averages $\chi_1^{[0]}(0)_0$ and  $\chi_1^{[1]}(0)_0$ are obtained 
by setting $J_0=0$ in Eqs.\ (\ref{chi1r}) and (\ref{chi1[1]}), respectively, 
and $\chi_1^{[1]}(0)$ is given by Eq.\ (\ref{chi1[1]}) with $y=0$.

We will also need the $n=2$ average 
\begin{equation}
\chi_1^{[2]}(0)_0 = {1\over 8\beta J^2} \;
 \left[{z^2\over r^3} - {1\over r}\right] \; . 
\label{chi1[2]}
\end{equation}

A numerical solution $z(T)$ of Eq.\ (\ref{eqzf}) in zero field ($E=0$)
can be found at all temperatures and is independent of $J_0$ as long
as $|J_0| < J$. An example is shown on Fig.\ 1 for $\Delta /J^2 = 0.001$. 
The inset shows that $z - \beta J$ 
is always positive, and since $2J$ is the largest eigenvalue of $J_{\lambda}$  
one can see that indeed $2z > 2\beta J_{\lambda}$ for all $\lambda$ as assumed 
earlier. When both $E \ne 0$ and $J_0 \ne 0$, there are in general two complex 
solutions for $z(E,T)$ and the present theory is not applicable. In the 
following we will only consider the cases in which a real solution $z(T)$ 
exists and has a real second derivative $z'' = d^2z/dE^2$ at $E=0$. 

For $\Delta = 0$ and $E=0$, Eq.\ (\ref{eqzf}) reduces to the equation derived 
in Ref. \cite{KTJ}, which has the solution $z_0 = (1+\beta^2J^2)/2$ for 
$T\ge J$. For $T < J$, however, the solution does not exist and $z$ must be 
obtained from the saddle-point condition \cite{KTJ}, yielding $z_0 = \beta J$. 

A numerical evaluation shows that the expression (\ref{chi1r}) for the static 
linear susceptibility fully agrees with the 
result obtained by means of replica theory in Ref.\ \cite{PB}.

One can also calculate the static third-order nonlinear susceptibility
$\chi_3$, which is defined in terms of the expansion
\begin{equation}
P = \chi_1 E - \chi_3 E^3 + \cdots  \; .
\label{PE}
\end{equation}
Obviously, $\chi_3 = -(1/6)(\partial^3P/\partial E^3)_{E=0}$. Using 
Eqs.\ (\ref{Pt}), (\ref{phia}) and (\ref{chi1[n]}) we find
\begin{equation}
\chi_3 = \chi_1^{[1]}(0) z_0'' \; .
\label{chi3}
\end{equation}
Evaluating $z_0'' \equiv (d^2z/dE^2)_{E=0}$ from Eq.\ (\ref{eqzf}) we obtain
the result
\begin{equation}
\chi_3 = \beta^2 {\chi_1^{[1]}(0)^2\over 
\chi_1^{[1]}(0)_0+2\beta^2\Delta \chi_1^{[2]}(0)_0} \; ,
\label{chi3s}
\end{equation}
where $\chi_1^{[2]}(0)_0$ is given by Eq.\ (\ref{chi1[2]}) above. The last expression has been 
evaluated numerically and found to be precisely equivalent to the result 
derived in Ref.\ \cite{PB} using the replica formalism.


\section{Dynamic response}

We now consider the case of an oscillating electric field 
$E(t) = E_0 \cos (\omega t)$. This is inserted into Eq.\ (\ref{Pt}).
At asymptotic times $t \gg t_c$ the response can be written by analogy 
with Eq.\ (\ref{PE}) as \cite{I1}
\begin{equation}
P(t) \sim \left[P_{\omega} e^{-i\omega t}
+P_{3\omega}e^{-i3\omega t} + \cdots \right] + {\mathrm c.c.} \; ,
\label{Pas}
\end{equation}
where $P_{\omega}$ and $P_{3\omega}$ are the amplitudes of the first 
and third harmonic response, respectively, which are given by
\begin{equation}
P_{\omega} = \chi_{1,0}(\omega)\left({E_0\over 2}\right) + 
\chi_{1,1}(\omega)\left({E_0\over 2}\right)^3 + \cdots \; ;
\label{P1}
\end{equation}
\begin{equation}
P_{3\omega} = \chi_{3,0}(\omega) \left({E_0\over 2}\right)^3
+\chi_{3,1}(\omega)\left({E_0\over 2}\right)^5 + \cdots \; .
\label{P3}
\end{equation}
Here we have introduced the linear dynamic response 
$\chi_{1,0}(\omega)$, the third-order nonlinear responses $\chi_{1,1}(\omega)$ 
and $\chi_{3,0}(\omega)$, etc. We will focus on the first harmonic linear 
response $\chi_{1,0}(\omega)$, which is equivalent to the dynamic 
linear susceptibility $\chi_1(\omega) = \chi_{1,0}(\omega)$, and on the
third-order nonlinear response $\chi_{3,0}(\omega)$. The latter is 
typically measured by monitoring the third harmonic component of $P(t)$ 
at small amplitudes of the field $E_0$ \cite{L1}. In order to ensure 
the proper static limit $\omega \to 0$ we will define the third-order
nonlinear dynamic susceptibility as $\chi_3(\omega) = - \chi_{3,0}(\omega)$.
From Eqs.\ (\ref{Pas})-(\ref{P3}) we thus find
\begin{equation}
\chi_1(\omega) = {\partial P_{\omega}  
\over \partial (E_0/2)}\Biggl\vert_{E_0=0} \; ;
\label{chi1d1}
\end{equation}
\begin{equation}
\chi_3(\omega) = -{1\over 6}{\partial^3 P_{3\omega} 
\over \partial (E_0/2)^3}\Biggl\vert_{E_0=0} \; .
\label{chi3d1}
\end{equation}

In the asymptotic regime, the function $\phi_{\lambda}(t)$ in 
Eq.\ (\ref{Pt}) behaves as
\begin{equation}
\phi_{\lambda}(t) \sim e^{-g_{\lambda}t - 2\varphi(t)} \; ,
\label{phias}
\end{equation}
where we have defined $\varphi (t) \equiv \int_0^tdt'[z(t')-z]$,
with $z$ representing the solution of Eq.\ (\ref{eqz}). The first part of the 
response (\ref{Pt}), which will be proportional to $\sim E_0\exp(-i\omega t)$, 
is now given by
\begin{equation}
P(t) = \beta \left({E_0\over 2}\right) \int_0^t dt_1 \ldb 
e^{-g_{\lambda}(t-t_1)}\rdb
e^{-2[\varphi(t)-\varphi(t_1)]} e^{-i\omega t_1}  + {\mathrm c.c.} 
\label{P-}
\end{equation}
%

\subsection{Linear dynamic susceptibility}

The part of $P_1(\omega)$, which is linear in $E_0$, is trivially obtained 
from Eq.\ (\ref{P-}) by noting that $\varphi(t)=0$ for $E_0=0$. We can 
thus evaluate the integral and using Eq.\ (\ref{chi1d1}) we find
\begin{equation}
\chi_1(\omega) = \beta \ldb {1\over g_{\lambda} - i\omega}
\rdb \; .
\label{chi1d2}
\end{equation}
Comparing with Eqs.\ (\ref{chi1s}) and (\ref{chi1[n]}) we realize that
the averages of the above type can be evaluated with the aid of 
Eq.\ (\ref{chi1[n]}), in which we set $y = \omega$ and $n=0$, yielding 
(with $\tau$ restored)
\begin{equation}
\chi_1(\omega) =  {z-i\omega \tau /2 - 
\sqrt{(z-i\omega \tau /2)^2 -\beta^2J^2}-\beta J_0\over 
\beta (J^2+J_0^2)-2 J_0(z-i\omega \tau /2)} \; .
\label{chi1d3}
\end{equation}
For $\omega \to 0$ this obviously reduces to the static susceptibility
(\ref{chi1r}).

The temperature behavior of $\chi_1(\omega)$ will crucially depend on
the temperature variation of the relaxation time $\tau (T)$. The SRBRF model
(\ref{H}) and the equations of motion (\ref{em1}) contain no information
about $\tau (T)$. It has been found empirically \cite{S1,VR1,VR2} that some of 
the properties of relaxors can be described by assuming a Vogel-Fulcher (VF)
relationship for $\tau$, namely,
\begin{equation}
\tau = \tau_0 \, \exp\left({U\over T-T_0}\right) \; , 
\label{VF}
\end{equation}
where $T_0$ is the VF temperature. This expression is valid for $T > T_0$ 
and would lead to a divergence of $\tau$ for $T \to T_0$. There is no 
{\it a priori} relation between $T_0$ and the parameters of the SRBRF model. 
A similar situation occurs in Ising dipolar glasses, where a probability
distribution of relaxation times $g(\ln \tau)$ has been used in combination 
with an empirical Debye-type response \cite{K1}. With $\tau$ lying in 
the range $\tau_{min} < \tau < \tau_{max}$, the VF temperature $T_0$ has
been identified with the freezing temperature $T_f$. On the other hand,
$\tau_{min}$ has been fitted to an Arrhenius-type expression 
$\tau_{min} \propto \exp (E/T)$. The same approach was found to
be applicable to relaxors as well \cite{L1}. 

An alternative approach is based on the master equation for the 
reorientation of cluster polarization assuming a VF relaxation time of the 
type (\ref{VF}), where the barrier heights $U$ are distributed according 
to a Gaussian probability function \cite{VR1,VR2}. Such an approach was 
found to be applicable to PMN and PST in the region $T > T_0$.

In general, we can thus introduce the average dynamic susceptibility
\begin{equation}
\overline{\chi}_1(\omega) = \int_{\tau_{min}}^{\tau_{max}}
{d\tau\over \tau}\,  g(\ln \tau) \chi_1(\omega) \; ,
\label{chiav}
\end{equation}
where the probability distribution of relaxation times $g(\ln \tau)$
is physically justified by the fact that relaxors are inherently inhomogeneous 
systems due to compositional disorder. Thus one may imagine, for example, 
that the relaxor system consists of a set of macroscopic regions, which
are formally characterized by the same microscopic equation of motion,
but differ in the value of the parameter $\tau$. 

One encounters serious difficulties in attempting to describe the 
dynamic response at $T < T_0$. Formally one could assume that 
$\tau \to \infty$ for $T \le T_0$, but this will lead to a zero value 
of $\chi_1(\omega) = 0$ at all temperatures $T \le T_0$. We can single out 
the following representative cases: (i) a single VF relaxation time (\ref{VF}); 
(ii) a nonsingular distribution of barrier heights $g(U)$; (iii) a distribution
of relaxation times $g(\ln \tau)$ such that its normalization 
$\int_{\tau_{min}}^{\tau_{max}} d\tau \, g(\ln \tau)/\tau$
diverges as $\tau_{max} \to \infty$. The first case is 
illustrated in Fig.\ 2, where we show the calculated real and imaginary 
parts of $\chi_1(T,\omega) = 0$ for several values of frequency 
assuming a single VF relaxation time (\ref{VF}). As in Fig.\ 1 we assume
$J_0 = 0.9 J $ and $\Delta /J^2 = 0.001$, as well as $T_0 = J$. Such behavior 
is incompatible with experiments, which generally show a smooth decrease 
of $\chi_1'(T,\omega)$ and $\chi_1''(T,\omega)$ across the region 
where $T_0$ is expected to be located. 

A more realistic description can be obtained, for example, by 
assuming a distribution $w(T_0)$ of VF temperatures $T_0$, where $T_0$ 
is allowed to vary in the range $0 < T_0 < T_0^{max}$. Using the 
relation $d(\ln \tau)g(\ln \tau) = dT_0 w(T_0)$ in Eq.\ (\ref{chiav}) 
and choosing a linear distribution $w(T_0) = 2(1-T_0/J)/J$ with $T_0^{max} = J$,
we obtain the temperature dependence of the linear susceptibility 
shown in Fig.\ 3. Here we used the same set of parameters as 
in Fig.\ 2. In contrast to the single VF temperature case, the 
above distribution leads to nonzero values of $\overline{\chi}_1(\omega)$ 
at all temperatures. The shape of the real and imaginary part of
$\overline{\chi}_1(\omega)$ is in qualitative agreement with the observed 
relaxation spectra in PMN \cite{L1} and PLZT \cite{K2}. 

It should be noted, however, that the above result for the linear 
susceptibility contains only the contribution of polar clusters. Other 
contributions may exist---for example, that of optic phonons---which are not
expected to show
any anomalies near $T_f$. In general, such contributions can be written
as a sum of Debye-like terms, with the possibility of an average
over the corresponding relaxation times.  
At present, the problem of a realistic relaxation time distribution in 
relaxors, which would be appropriate at all temperatures, has not yet
been resolved.

\subsection{Zero-field-cooled susceptibility}

In analogy with Ising spin glasses \cite{SZ} and dipolar glasses \cite{TPB}
one can introduce an effective relaxation time for the low-frequency response
\begin{equation}
\tau_{eff} = -i\beta \,\lim_{\omega \to 0} {\partial
 \chi_1(\omega)^{-1}\over \partial \omega} \; .
\label{taueff}
\end{equation}
Returning to Eq.\ (\ref{chi1d1}) and using the above definition of 
$\chi_1^{[1]}(y)$ from Eq.\ (\ref{chi1[1]}) in which we set 
$y=\omega\tau$, we obtain the result
\begin{equation}
\tau_{eff} = {1\over 2}\beta\, \tau\, \lim_{\omega \to 0}
{\chi_1^{[1]}(\omega\tau)\over \chi_1(\omega)^2} \; .
\label{tau1}
\end{equation}

It will be shown later that the real part of the nonlinear susceptibility 
$\chi_3(\omega)$ is also proportional to $\chi_1^{[1]}(\omega)$ and
in the static limit shows a sharp peak at the "freezing" temperature
$T_f \approx J$ for $\Delta \ll J^2$, while it actually diverges if 
$\Delta = 0$. Since $\chi_1(\omega)$ is a well behaved function of 
temperature, it follows that $\tau_{eff}$ increases as $\sim (T-J)^{-1}$ 
on approaching $T_f$, however, it remains finite at $T_f$. Thus the 
behavior of $\tau_{eff}$ mimics the freezing transition in this case.
A true freezing transition at $T_f = T_0$ can be described by 
assuming a VF temperature dependence of $\tau$, which is then transferred 
to $\tau_{eff}$ via Eq.\ (\ref{tau1}).

When $\tau_{eff}$ is large, the system will only reach equilibrium 
at times $t \gg \tau_{eff}$, which may become very long. In measuring the  
static susceptibility $\chi_1$ one should distinguish between the
field-cooled and zero-field-cooled susceptibility, $\chi_1^{FC}$ and
$\chi_1^{ZFC}$, respectively. Here $\chi_1^{FC}$ is given by Eq.\ (\ref{chi1s}) 
and corresponds to an experiment carried out on a time scale 
$t \gg \tau_{eff}$. On the other hand, $\chi_1^{ZFC}$ can be obtained 
by turning on a field $E(t_1) = E$ at time $t_1=0$ and measuring the
response $P(t)$ at time $t > t_1$ as described by Eq.\ (\ref{Pt}). Thus
we can write
\begin{equation}
\chi_1^{ZFC}(t) = {\partial P(t)\over \partial E} 
= \beta\int_0^t dt_1  \ldb{\phi_{\lambda}(t)\over
\phi_{\lambda}(t_1)}\rdb \; ,
\label{chizfc1}
\end{equation}
and using Eq.\ (\ref{phia}) we find
\begin{equation}
\chi_1^{ZFC}(t) = \beta\ldb{1 - e^{-(2z-\beta J_{\lambda})t/\tau}
\over 2z-\beta J_{\lambda}}\rdb \; .
\label{chizfc2}
\end{equation}
For $t \to \infty$ this reduces to the previous result (\ref{chi1s}),
which corresponds to $\chi_1^{FC}$. The difference between the two 
susceptibilities $\delta\chi_1(t) \equiv \chi_1^{FC} - \chi_1^{ZFC}(t)$ 
is given by

\begin{equation}
\delta\chi_1(t) = \beta\ldb{e^{-(2z-\beta J_{\lambda})t/\tau}
\over 2z-\beta J_{\lambda}}\rdb
= \beta\int_{t/\tau}^{\infty} dt_1\ldb 
e^{-(2z-\beta J_{\lambda})t_1}\rdb \; .
\label{dchi1}
\end{equation}
The value of the integral can be estimated by first using Eqs.\ (\ref{av})
and (\ref{rho}) and expanding the integrand in powers of $J_0/J$ \cite{RPu}.
To leading order the result is independent of the parameter $J_0$ and shows 
that for $\Delta \ll J^2$ and $T \le J$ one has a power law behavior 
$\delta\chi_1(t) \sim (t/\tau )^{-1/2}$, implying a slow decay and a 
large difference between the two susceptibilities, which has been
observed experimentally \cite{L1}. On the other hand, for $T > J$ the 
asymptotic behavior is a combination of power law and exponential, i.e., 
$\delta\chi_1(t) \sim (t/\tau )^{-3/2}\exp[-2(z-\beta J)t/\tau]$. Thus
in this regime the difference decays much faster and the two susceptibilities 
become indistinguishable on a typical experimental time scale. 


\section{Nonlinear dielectric susceptibility}

To calculate the third order partial derivative in Eq.\ (\ref{chi3d1}) 
we return to Eq.\ (\ref{P-}), in which $\varphi (t)$ is now
a function of $E_0$. In general, $\varphi(t)$ will be a sum of terms,
which are even powers of $E_0e^{\pm i\omega t}$. We will focus on the 
second-order term $ \sim E_0^2e^{-2i\omega t}$. Introducing the function
\begin{equation}
X(t) = {\partial^2\varphi (t)\over \partial (E_0/2)^2}
\Biggl\vert_{E_0=0} \; 
\label{Xt}
\end{equation}
we can express the third partial derivative in Eq.\ (\ref{chi3d1}) as
\begin{equation}
{\partial^3P_3(\omega )\over \partial (E_0/2)^3}\Biggl\vert_{E_0=0} = 
-6\beta e^{2i\omega t}\int_0^t
dt_1 e^{i\omega (t-t_1)}\ldb e^{-g_{\lambda}(t-t_1)}\rdb
[X(t)-X(t_1)]  \; .
\label{P3E3}
\end{equation}
The function $X(t)$ will be calculated from Eq.\ (\ref{ctt}) in the 
asymptotic limit. Considering only the terms, which asymptotically behave 
as $\sim e^{-2i\omega t}$ and taking the second derivative with respect
to $E_0/2$ leads to
\begin{eqnarray}
0=&&-4\left[\int_0^tdt_1 \ldss e^{-2g_{\lambda}(t-t_1)}\rdss_{\!0}
-2\beta^2\Delta \int_0^tdt_1 \int_0^tdt_2\lds
e^{-2g_{\lambda}(2t-t_1)-t_2}\rds_{\!0}\right][X(t)-X(t_1)]\nonumber\\
&&+\beta^2\int_0^tdt_1 \int_0^tdt_2\lds
e^{-2g_{\lambda}(2t-t_1-t_2}\rds e^{-i\omega (t_1+t_2)} \; .
\label{cttX}
\end{eqnarray}
The last double integral becomes for asymptotic values of $t$ 
\begin{equation}
+\beta^2 e^{-2i\omega t}\ldb {1\over 
\;(g_{\lambda}-i\omega)^2}\rdb \; .
\label{int2}
\end{equation}

We now apply the Laplace transform to Eq.\ (\ref{cttX}) using the definition
\begin{equation}
{\tilde X}(p) = \int_0^{\infty} dt \, e^{-pt} X(t) \;  
\label{LT}
\end{equation}
and obtain the result:
\begin{equation}
{\tilde X}(p) = {\beta^2\over 2(p+2i\omega )}\; 
{\lds {1\over (g_{\lambda}-i\omega)^2}\rds 
\over \lds {1\over g_{\lambda}}\rds_{\!0}
- \lds {1\over g_{\lambda}+p/2}\rds_{\!0}
+\beta^2\Delta\left[\lds {1\over g_{\lambda}^2}\rds_{\!0}
-\lds {1\over g_{\lambda}(g_{\lambda}+p)}\rds_{\!0}\right]} \; .
\label{Xp}
\end{equation}
The averages can be expressed in terms of the generalized responses 
(\ref{chi1[1]}). 

The function $X(t)$ can  be obtained by the inverse Laplace transform. 
Its behavior is determined by the poles $p_k = p'_k+ip''_k$ of 
${\tilde X}(p)$. A numerical evaluation shows that all poles 
are such that $p'_k \le 0$. At asymptotic times only those poles
for which $p'_k=0$ will be relevant. There is only one such pole, namely,
$p_0=-2i\omega$ leading to
\begin{equation}
X(t) \sim \left({\beta^2\over 2}\right) {\chi_1^{[1]}(\omega)\, e^{-2i\omega t}
\over \chi_1(0)_0 - \chi_1(\omega)_0+\beta^2\Delta \left[\chi_1^{[1]}(0)_0 
+{\chi_1(0)_0-\chi_1(2\omega )_0\over 2i\omega} \right]} \; .
\label{Xta}
\end{equation}

Inserting this expression into Eq.\ (\ref{P3E3}), evaluating the integral, and
returning to Eq.\ (\ref{chi3d1}), we obtain the final result for the complex
third-order nonlinear susceptibility (with $\tau$ restored):
\begin{equation}
\chi_3(\omega) = \left({\beta^2\over 2}\right) {\chi_1^{[1]}(\omega\tau) \,
[\chi_1(\omega)-\chi_1(3\omega)] 
\over \chi_1(0)_0 - \chi_1(\omega)_0+\beta^2\Delta \left[\chi_1^{[1]}(0)_0 
+{\chi_1(0)_0-\chi_1(2\omega )_0\over 2i\omega\tau} \right]} \; .
\label{chi3f}
\end{equation}
Here $\chi_1^{[1]}(\omega\tau)$ is given by Eq.\ (\ref{chi1[1]}) with
$y=\omega\tau$, and  $\chi_1(\omega)$ by Eq.\ (\ref{chi1d3}).

This expression may now be averaged over the distribution of relaxation 
times $\tau$ as argued in the paragraph preceding Eq.\ (\ref{taueff}).
In Fig.\ 4  the real and imaginary parts of $\overline{\chi}_3(\omega)$ are plotted 
as functions of temperature for several values of frequency. In analogy with 
the case of $\overline{\chi}_1(\omega)$ above, a VF behavior (\ref{VF}) of $\tau$ 
and a linear distribution of VF temperatures $T_0$ has been used. The values of 
the parameters are again $J_0/J = 0.9$ and $\Delta /J^2 = 0.001$. The real 
part $\overline{\chi}_1'(T,\omega)$ has a sharp peak near $T \simeq J$, whose
 origin can be traced back to the function $\chi_1^{[1]}(\omega)$ appearing in 
Eq.\ (\ref{chi3f}). As in the linear susceptibility case, a strong
frequency dispersion is evident. In the limit of small frequencies, i.e., 
$\omega \tau_0 \ll 1$, the behavior of $\overline{\chi}_3'(T)$ is the same 
as in the static case studied by replica theory \cite{PB}. At high 
frequencies and low temperatures, $\overline{\chi}_3'(T,\omega)$ may become 
negative due to the last factor in the numerator of Eq.\ (\ref{chi3f}), whose 
imaginary part changes sign. This effect cannot be observed easily, since this
would require a measurement of the nonlinear susceptibility in the range 
where the absolute value of $\overline{\chi}_3'(T,\omega)$ is extremely small
compared to its peak value. 

It is easily verified that in the limit $\omega \to 0$, Eq.\ (\ref{chi3f})
reduces to the static result (\ref{chi3s}).

It has been shown in the static theory \cite{PB} that a crucial quantity,
which can discriminate between the dipolar glass-like and ferroelectric 
behavior, is not $\chi_3(T)$ but rather the rescaled static nonlinear 
susceptibility $a_3 = \chi_3/\chi_1^4$. In spin glasses without random 
fields $a_3(T)$ diverges at $T_f$, and in a relaxor it develops a peak 
near $T_f$, whereas in a ferroelectric with long range order $a_3$ decreases 
with decreasing temperature on approaching the critical temperature 
$T_c$ \cite{L1}. It has been suggested that $a_3(T)$ could also be
extracted from the dynamic linear and nonlinear susceptibilities by
considering the following generalized function \cite{I1,G1}:
\begin{equation}
a_3'(T,\omega) = {\overline{\chi}_3'(\omega)\over 
\overline{\chi}_1'(3\omega)\overline{\chi}_1'(\omega)^3} \; .
\label{a3}
\end{equation}

In Fig.\ 5, $a_3'(T,\omega)$ is  plotted as a function of temperature 
for the same set of parameters as in Fig.\ 4. Each of the three factors 
in Eq.\ (\ref{a3}) has been averaged over the linear distribution of $T_0$. 
Obviously, $a_3'(T,\omega)$ develops a peak near $T_f \simeq J$ at all 
frequencies shown. On the high-$T$ side of the peak, $a_3'(T)$ is 
independent of frequency and agrees with the static $a_3$. Near the peak and
on its low-$T$ side, however, strong frequency dispersion appears. Recent
experiments in PMN and PLZT \cite{B5} indicate that the high-$T$ quasistatic
part of $a_3(T)$ exhibits a crossover between the  
paraelectric-like decreasing behavior and a glass-like increasing behavior 
on approaching $T_f$. This type of behavior is characteristic of relaxor 
ferroelectrics. The crossover can be qualitatively described by the present 
dynamic SRBRF model, as shown in the inset of Fig.\ 5. The model also correctly
predicts the onset of strong frequency dispersion in $a_3'(T,\omega)$ 
at low temperatures, which has been observed in both PMN and PLZT \cite{B5}.

\section{Conclusions}

We have presented a dynamic model of uniaxial relaxor ferroelectrics based 
on the recently developed static spherical random-bond---random-field (SRBRF)
model \cite{PB}. Following the theory of spherical models of spin 
glasses \cite{CD1,CD2} the order parameter field is assumed to obey the
Langevin equation of motion written in the representation of eigenstates
of the random interaction matrix with the spherical condition being enforced 
at all times. The equations of motion
are exactly solvable in the asymptotic limit where the relaxor system 
reaches an equilibrium state. The linear and the third-order nonlinear dynamic
response functions have been derived. In the static $\omega \to 0$ limit these
results are precisely equivalent to the static linear and nonlinear 
susceptibilities $\chi_1$ and $\chi_3$, respectively, obtained earlier by
the replica method \cite{PB}. 

In analogy with the static case, the dynamic theory does not predict a sharp 
transition into a dipolar glass-like state. Rather, in the case of weak 
random fields the third-order susceptibility shows a narrow peak at a 
temperature $T_f$, which mimics the freezing transition. Within the context
of a dynamic model the freezing transition would correspond to the divergence
of the longest relaxation time in the system. However, the dynamic SRBRF model 
contains no information on the behavior of the relaxation time $\tau$ 
appearing in the equations of motion, and does not lead to the divergence 
of the effective relaxation time on approaching $T_f$. In order to describe
the observed freezing transition one should therefore introduce a divergent behavior
of $\tau$. This can be done, for example, by assuming a Vogel-Fulcher (VF) law 
for the temperature behavior of $\tau$ in accordance with empirical findings 
\cite{VR1,VR2}, however, this will suppress the dynamic response at all 
temperatures lower than the VF temperature $T_0$. We have shown that by
introducing a probability distribution of VF temperatures $T_0$ one can obtain
linear and nonlinear response functions which remain finite at all temperatures,
in qualitative agreement with experiments. The largest value of $T_0$ has
been set equal to the static "freezing" temperature $T_f$, which is 
determined by the random bond strength parameter $J$.

The actual shape of $\overline{\chi}_1(T,\omega)$ and $\overline{\chi}_3(T,\omega)$,
where the bar means an average over $\tau$ or equivalently over $T_0$, strongly 
depends on the probability distribution of relaxation times $g(\ln \tau)$.

Within the framework of the dynamic SRBRF model we have also calculated 
the scaled dynamic nonlinear susceptibility 
$a_3'(T,\omega) = \overline{\chi}_3'(\omega)/
\overline{\chi}_1'(3\omega)\overline{\chi}_1'(\omega)^3$,
which allows one to discriminate between the ferroelectric-like and
glass-like behavior of relaxors \cite{B5,G1}. In the quasistatic regime 
above $T_f$, $a_3'(T,\omega)$ is practically independent of $\omega$ and its 
temperature dependence shows a crossover between paraelectric-like and 
glass-like behavior on approaching $T_f$ from above. This crossover behavior 
has recently been observed both in PMN and PLZT \cite{B5}. 
The calculated shape of $\overline{\chi}_1(T,\omega)$ and 
$\overline{\chi}_3(T,\omega)$, and hence of $a_3'(T,\omega)$, 
strongly depends on the probability distribution of relaxation times $g(\ln\tau)$. 
In the present work we did not attempt to investigate in detail the effects 
of $g(\ln \tau)$ on the behavior of these quantities; however, we have 
shown that if one assumes a linear distribution of VF temperatures $T_0$, 
the predicted behavior of $\overline{\chi}_1(T,\omega)$ and 
$\overline{\chi}_3(T,\omega)$ is in qualitative agreement with experiments 
in PMN \cite{B5,L1,G1} and PLZT \cite{K2,B5}. On the other hand, 
$a_3'(T,\omega)$ obtained in this manner has a peak near
$T_f \simeq J$ and shows a strong frequency dispersion below $T_f$. The
predicted peak in $a_3'(T,\omega)$ has not been observed 
experimentally \cite{G1}, suggesting that one should perhaps search for a more 
realistic distribution $g(\ln\tau)$. This problem will be dealt with in a 
future publication.

\acknowledgements{}

This work was supported by the Ministry for Science and Technology of the
Republic of Slovenia. One of the authors (R.\ P.) is grateful to Silvio 
Franz and to Boris E. Vugmeister for helpful suggestions.


\begin{figure}[h]
\caption{Solution of Eq.\ (23) for the Lagrange multiplier 
$z(T)$ obtained numerically in zero field ($E=0$) and for $\Delta/J^2 = 0.001$. 
Inset: Proof that $z - \beta J > 0$ at all temperatures.}
\label{fig1}
\end{figure}
\begin{figure}[h]
\caption{Real and imaginary parts of the linear susceptibility in the case
of a single Vogel-Fulcher (VF) type relaxation time [Eq.\ (37)] as functions 
of temperature for several values of frequency, as indicated. Note that
the response is strictly zero below the VF temperature $T_0 = J$.}
\label{fig2}
\end{figure}
\begin{figure}[h]
\caption{Real and imaginary parts of the linear susceptibility averaged 
over a linear probability distribution of VF temperature $T_0$, 
with $0 < T_0 < J$.}
\label{fig3}
\end{figure}
\begin{figure}[h]
\caption{Calculated temperature dependence of the real and imaginary parts
of the third-order nonlinear susceptibility averaged over a linear 
distribution of $T_0$, using the same parameter values as in Figs.\ 2 and 3.}
\label{fig4}
\end{figure}
\begin{figure}[h]
\caption{Temperature dependence of the scaled nonlinear susceptibility \\
$a_3'(T,\omega) = \overline{\chi}_3'(\omega)/
\overline{\chi}_1'(3\omega)\overline{\chi}_1'(\omega)^3$
obtained with the same parameter values as in Figs.\ 2-4. The inset shows
the crossover from the decreasing paraelectric-like to increasing 
glass-like behavior in the quasistatic regime above the peak.}
\label{fig5}
\end{figure}

\end{document}